\documentclass[%
 reprint,
 amsmath,amssymb,
 aps, longbibliography
]{revtex4-1}

\usepackage{graphicx}
\usepackage{dcolumn}
\usepackage{bm}
\usepackage{tipa}
\usepackage{upgreek}

\usepackage{amsmath}
\usepackage{bm}
\usepackage{xcolor}
\usepackage{soul}
\usepackage[normalem]{ulem}
\usepackage{siunitx}
\usepackage[version=4]{mhchem} 
\usepackage[capitalize]{cleveref}

\usepackage{xcolor, soul}
\sethlcolor{yellow}

\begin{document}

\title{Bipolar surface charging by evaporating water droplets}


\author{Nitish Singh$^{1}$}
\thanks{These authors contributed equally to this work.}
\author{Aaron D. Ratschow$^{2}$}
\thanks{These authors contributed equally to this work.}
\author{Nabeel Aslam$^{3}$}
\author{Dan Daniel$^{1,4}$}
    \email{dan-daniel@oist.jp}

\affiliation{$^{1}$Droplet Lab, Division of Physical Sciences and Engineering, King Abdullah University of Science and Technology (KAUST), Thuwal 23955-6900, Saudi Arabia}
\affiliation{$^{2}$Max Planck Institute for Polymer Research, 55128 Mainz, Germany}
\affiliation{$^{3}$Nanofabrication Core Lab, King Abdullah University of Science and Technology (KAUST), Thuwal 23955-6900, Saudi Arabia}
\affiliation{$^{4}$Droplet and Soft Matter Unit, Okinawa Institute of Science and Technology Graduate University, Onna, Okinawa 904-0495, Japan}
\begin{abstract}
Surface charging is a ubiquitous phenomenon with important consequences. On one hand, surface charging underpins emerging technologies such as triboelectric nanogenerators; on the other, uncontrolled charging can damage delicate nanostructures and devices. Despite its significance, surface charging by evaporating water droplets remains poorly understood. Here, using Kelvin Probe Force Microscopy, we spatially resolve the surface-charge patterns from evaporating droplets and propose a physical model that quantitatively explains the origin of bipolar charging. 
\end{abstract}

\maketitle

Surface charging by water droplets can occur through many routes---pipetting \cite{Choi.2013}, spraying \cite{Nolan.1922,Baumann.2021}, droplet impact \cite{Lenard.1892,Levin.1971,Diaz.2022,Piednoir.2025}, or even jumping droplets \cite{Miljkovic.2013,Miljkovic.2014}---each with wide-ranging technological relevance and practical consequences. For example, surface charging can generate electricity through triboelectric nanogenerators \cite{Kwon.2014,Wu.2020,Wang.2021,Xu.2020,Helseth.2021}. Conversely, the same phenomenon can trigger electrostatic discharges capable of igniting explosions \cite{anderson1978sparks} and damaging nanodevices \cite{swan2016impact}. More broadly, surface charging can influence droplet motion \cite{Li.2022,Sun.2019, Li.2023}, selectively deposit molecules \cite{Li2022surfactant,Zhou.2025}, and suppress splashing \cite{yu2025charged}.

One route that has become a focal point of intense research in recent years is slide electrification: a droplet sliding over an insulating substrate acquires an electric charge while depositing charge of the opposite polarity along its path \cite{Yatsuzuka.1994, Stetten.2019, Ratschow.2025}. 
The resulting surface-charge pattern is unipolar \cite{Coehn.1916,Sosa.2020,Nie.2020,Artemov.2023}, and its magnitude decays exponentially with sliding distance \cite{Stetten.2019,Ratschow.2025, Bista.2023}.

The origin of slide electrification, and surface charging in general, lies in charge transfer at the receding contact line, commonly attributed to ions in the electric double layer \cite{Langmuir.1938,Ratschow.2024, Bista.2024}. While electron transfer has been proposed as an alternative mechanism, it remains highly contested \cite{Lin.2020,Jin.2024}. Slide electrification is well-documented on hydrophobic surfaces, but less so on hydrophilic ones \cite{Ratschow.2025}.

In contrast to sliding droplets, surface charging by evaporating droplets remains largely unexplored \cite{Knorr.2024, He.2019}, and differences between these scenarios remain underappreciated. Notably, the droplet footprint shrinks during evaporation but remains constant during sliding. The contact-line velocity is also much lower in evaporating droplets ($\sim\,$\SI{}{\micro\metre\per\second}) than in sliding droplets ($\sim \,$\SI{}{\milli\metre\per\second} or more \cite{Hinduja.2024}); hence, hydrodynamic flow plays little role in evaporation-driven charge transfer \cite{Ratschow.2024}. While slide electrification yields unipolar charging, evaporating droplets produce bipolar surface-charge patterns \cite{He.2019,lin2023size, Knorr.2024, Moreira2020}. The mechanism behind bipolar charging remains elusive, with prior studies limited by low-resolution imaging and long acquisition times (up to hours \cite{He.2019}), during which surface charge may dissipate \cite{navarro2023surface,Bista.2024}. 

Here, we spatially map the surface-charge patterns from evaporating microdroplets across various substrates---from polymers to ceramics and from hydrophobic to hydrophilic surfaces---with micron resolution and within minutes.  In all cases, the charge patterns are bipolar: the droplet initially deposits charge of one polarity, which later reverses. We present an analytical model that quantitatively captures the observed patterns and, when fitted to experimental data, yields key properties of the electric double layer.

Our findings have potentially important implications: the ability to control surface-charge patterns enables site-specific deposition of colloids and molecules \cite{aizenberg2000patterned, thakur2013directed, Zhou.2025}, while avoiding unwanted charge can minimize dust and pollutant adhesion on windows and solar panels \cite{panat2022electrostatic}.

\begin{figure}[!htb]
\centering
\includegraphics[scale=0.95]{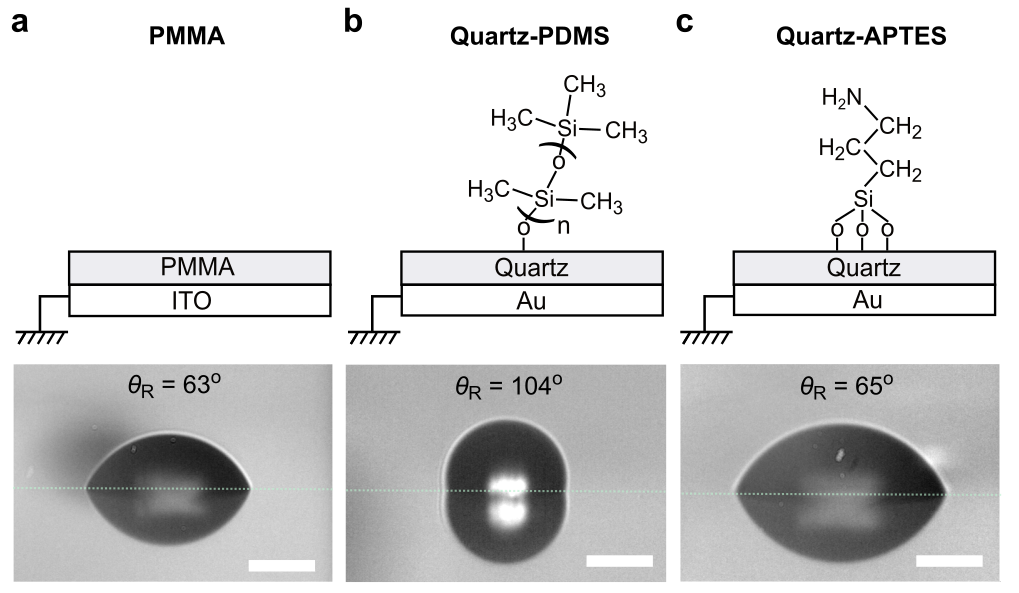}
\caption{\label{fig:surfaces} Three surfaces studied: (a) PMMA, (b) Quartz-PDMS, and (c) Quartz-APTES. \textit{Bottom:} side-view images during retraction. All scale bars are \SI{50}{\micro\meter}.}
\end{figure}

\emph{Experimental approach.} We initially investigated charging on three substrates [\cref{fig:surfaces}]: a \SI{4}{\micro\metre}-thick film of bare polymethylmethacrylate (PMMA); a \SI{0.4}{\micro\metre}-thick amorphous quartz film with grafted-to polydimethylsiloxane brushes (Quartz-PDMS) \cite{wang2016covalently}; and an identical quartz functionalized with a self-assembled monolayer of 3-aminopropyl-triethoxysilane (Quartz-APTES). All are supported on ITO or gold-coated glass which is grounded [details in Supplemental Material (SM)]. Quartz-PDMS is hydrophobic with a receding contact angle $\theta_{r} = $ \ang{104}, whereas PMMA and Quartz-APTES are hydrophilic, with $\theta_{r} = $ \ang{63} and \ang{65}, respectively. All substrates are highly insulating dielectrics and retain charges for over \SI{15}{\minute} [SM Fig.~S1].

\begin{figure}[!htb]
\centering
\includegraphics[scale=0.95]{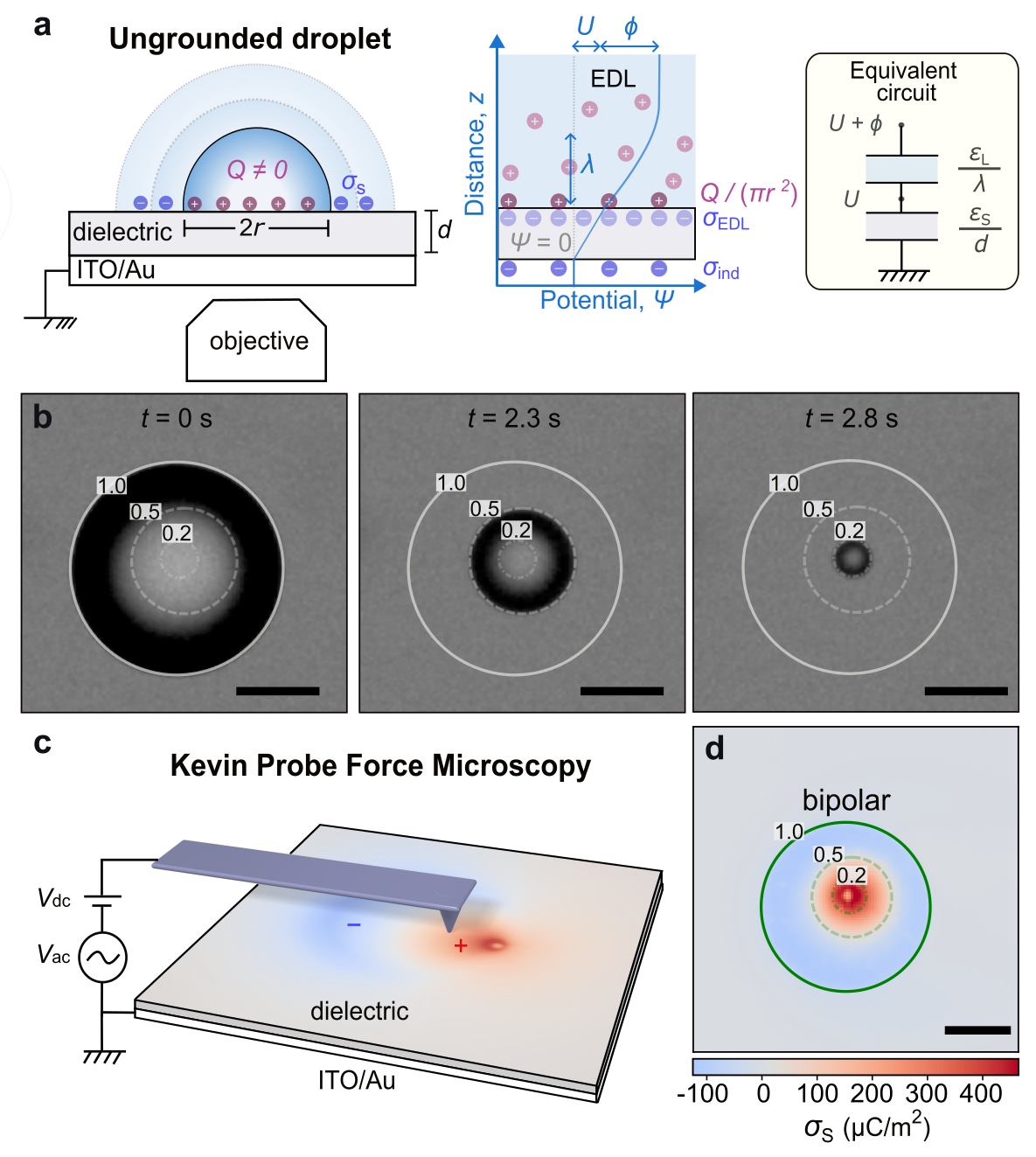}
\caption{\label{fig:mapping} (a) \textit{Left:} Contact-line positions tracked optically. \textit{Right:} Schematic of the electric double layer (EDL) and the equivalent circuit. (b) Time-lapse images on Quartz-PDMS. Labels indicate $r/r_0$. (c) Kelvin Probe Force Microscopy used to map the surface-charge patterns. (d) Bipolar pattern for (b).  All scale bars are \SI{20}{\micro\meter}.}
\end{figure}

Deionized water microdroplets were generated using a plastic spray bottle. The resulting droplets were polydisperse, with initial footprint radii $r_0 = 29\pm7 \,$\SI{}{\micro \meter}, and carried mimimal charges of $|Q_{0}| < 80\,\text{fC}$. During evaporation, the contact-line positions were recorded with an inverted optical microscope [Fig.~\ref{fig:mapping}a]. Under ambient conditions (\SI{22}{\celsius}, 50\% relative humidity), the droplet lifetime was on the order of seconds \cite{picknett1977evaporation}, with a typical contact-line speed of \(u \sim \SI{10}{\micro\metre\per\second}\) [Fig.~\ref{fig:mapping}b]. This translates to a P{\'e}clet number Pe $= u \lambda/D \approx 2 \times 10^{-3}$, assuming a Debye length $\lambda \approx 200$ nm for pure water and an ionic diffusivity $D \approx 10^{-9}$ m$^{2}$ s$^{-1}$. Since Pe $\ll 1$, the charge transfer process is quasi-static and not influenced by hydrodynamic flow \cite{Ratschow.2024}.  

Following evaporation, we used Kelvin Probe Force Microscopy (KPFM) to spatially map the resulting surface-charge pattern $\sigma_{\text{S}}(x, y)$ with micron resolution [Fig.~\ref{fig:mapping}c]. The acquisition time was 2 min, much faster than the charge dissipation timescale. A droplet on Quartz-PDMS [Fig.~\ref{fig:mapping}b] with an initial $r_{0}$ = \SI{26}{\micro\metre} produced a bipolar surface-charge pattern [Fig.~\ref{fig:mapping}d]; the normalized radial coordinate  $r/r_{0}$ is supermiposed to show the contact line at different times. The droplet charge $Q$ at any one time can then be obtained by integrating $\sigma_{\text{S}}(x, y)$ within the area enclosed by $r(t)$: 
\begin{equation}
Q(r) = \iint_{A(r)} \sigma_{\mathrm{S}}(x, y) \, \mathrm{d}A,   \label{eq:charge_integral}
\end{equation}
with the initial charge given by $Q_{0} = Q(r_{0})$.

Initially at $r/r_{0} = 1$, the \textit{ungrounded} droplet deposits maximal negative $\sigma_{\mathrm{S}}$ = $-\SI{104}{\micro C \, m^{-2}}$ (blue), which reverses to positive $\sigma_{\mathrm{S}}$ = $+\SI{345}{\micro C \, m^{-2}}$ (red) as it shrinks to $r/r_{0} = 0.2$. Interestingly, there is a region of zero charge $\sigma_{\mathrm{S}} = 0$ (light grey) at \( r/r_0 = 0.5 \). This bipolar pattern is a hallmark of evaporation-driven charging, distinct from the unipolar patterns in slide electrification. It results from the evolving droplet charge $Q$ and changing droplet potential $U$ [Fig.~\ref{fig:mapping}a, right schematic]. 

\begin{figure}[!htb]
\centering
\includegraphics[scale=1]{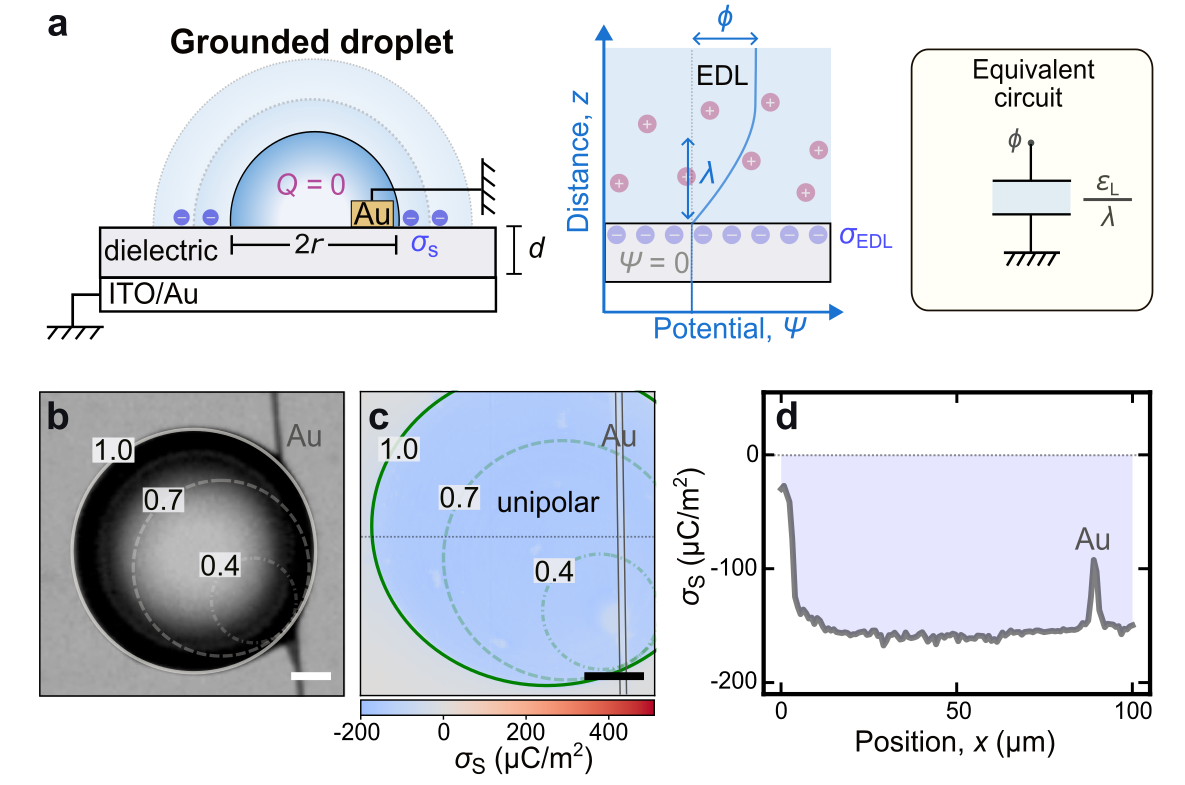}
\caption{\label{fig:unipolar} (a) \textit{Left:} An evaporating droplet contacting a grounded Au strip. \textit{Right:} Schematic of the EDL and the equivalent circuit. (b, c) The grounded droplet deposits unipolar pattern on Quartz-PDMS surface, (d) with uniform, negative $\sigma_{\mathrm{s}}$. All scale bars are \SI{20}{\micro\meter}.}
\end{figure}

\emph{Unipolar charging}. Bipolar surface charging can be completely suppressed by grounding. To this end, we first consider the case of a droplet contacting a grounded gold strip (width: \SI{3}{\micro \meter}, thickness: \SI{0.4}{\micro \meter}) [\cref{fig:unipolar}a, b]. In this control experiment, the droplet carries no charge ($Q = 0$, $U = 0$) and deposits a unipolar surface-charge pattern [\cref{fig:unipolar}c], with a constant negative $\sigma_{\mathrm{S},0} = -159\pm 3\,$\SI{}{\micro \coulomb}$\,$m$^{-2}$ on Quartz-PDMS [\cref{fig:unipolar}d].

The origin of surface charge is the electric double layer (EDL) that forms when a solid contacts water. The EDL consists of bound surface charges $\sigma_{\text{EDL}}$ and a diffuse counter-ion cloud that extends over the Debye length $\lambda$ [\cref{fig:unipolar}a]. As the contact line recedes, a fraction $\alpha < 1$ of $\sigma_{\text{EDL}}$ is transferred to the dewetted surface, giving  
\begin{equation}
    \sigma_\mathrm{\mathrm{S},0}=\alpha \sigma_\mathrm{EDL}=\alpha \left(\frac{K \varepsilon_{\text{L}} \phi}{\lambda}\right). \label{eq:sigma_s_grounded}
\end{equation}
Here, $\varepsilon_\text{L}$ is the liquid permittivity and $\phi$ is the potential drop across the EDL, i.e., the surface potential. The surface charge $\sigma_{\text{EDL}} \simeq \varepsilon_\text{L} \phi/\lambda$ follows from linearizing the Grahame equation under the Debye-H{\"u}ckel approximation, treating the EDL as a capacitor with areal capacitance $\varepsilon_\text{L}/\lambda$ [\cref{fig:unipolar}a]. The correction factor $K$ captures EDL distortion near the contact line due to the wedge geometry of the water–air interface. \cite{Dorr.2012,Dorr.2014,Ratschow.2024}. 

Since the charge transfer coefficient $\alpha$ is a constant for Pe $\ll$ 1 \cite{Ratschow.2024}, \cref{eq:sigma_s_grounded} predicts a uniform, unipolar charge pattern, consistent with the experimental observations [\cref{fig:unipolar}c, d]. 

\emph{Bipolar charging.} The previous model must be modified to account for charge accumulation for the more general case of ungrounded droplets. The change in droplet charge is given by the derivative of \cref{eq:charge_integral}
\begin{equation}
    \mathrm{d}Q=2\pi\sigma_\mathrm{S}r\mathrm{d}r. \label{eq:chargebalance}
\end{equation}
Since the evaporating droplet shrinks ($\mathrm{d}r < 0$), \cref{eq:chargebalance} implies that the droplet acquires charge with opposite polarity to $\sigma_\mathrm{S}$.

The resulting droplet potential is given by $U = Q/C$, where $C = \pi r^2 \varepsilon_\mathrm{S}/d$ is the capacitance for a dielectric of permittivity $\varepsilon_\mathrm{S}$ and thickness $d$. Charge transfer at the solid-liquid interface is therefore governed by two capacitors in series: (i) the EDL with areal capacitance $\varepsilon_{\mathrm{L}}/\lambda$ and (ii) the dielectric layer with areal capacitance $\varepsilon_{\mathrm{S}}/d$ [\cref{fig:mapping}a]. The surface charge deposited follows from Gauss's law and is given by
\begin{equation}
\begin{aligned}
	    \sigma_\mathrm{s} &=\alpha\left( \frac{K\varepsilon_\mathrm{L}\phi}{\lambda} + \frac{\varepsilon_\mathrm{S}U}{d} \right) \\
                          &= \alpha \left( \sigma_\mathrm{EDL} + \frac{Q}{\pi r^{2}}\right),
 \label{eq:sigma_s_of_U}	
\end{aligned}
\end{equation}
which reduces to \cref{eq:sigma_s_grounded} in the grounded case $U$ = 0. We assume that $\phi$ does not change with $U$, consistent with prior slide electrification experiments \cite{Bista.2023,Ratschow.2025}. 

Physically, $\sigma_\mathrm{s}$ reflects a balance between two opposing effects: $\phi$, which promotes charge deposition of one polarity, and $U$, which drives charge of the opposite polarity. During evaporation, the droplet footprint shrinks, decreasing the capacitance $C$, while charge accumulation causes $Q$ to increase. These two effects act synergistically to raise $U$ until it exceeds a threshold voltage $U^*=-K\phi\varepsilon_\mathrm{L}d/(\varepsilon_\mathrm{S} \lambda)$, triggering a polarity reversal. In contrast, during slide electrification, the footprint and capacitance remain constant, and the potential saturates at $U^*$; thus, while $\sigma_\mathrm{s}$ may decay to zero, no reversal occurs.

Substituting \cref{eq:sigma_s_of_U} into \cref{eq:chargebalance} yields an ordinary differential equation for $Q$
\begin{equation}
    \frac{\mathrm{d}Q}{\mathrm{d}r}=2\alpha \left( \pi r \sigma_\mathrm{EDL} + \frac{Q}{r}\right), \label{eq:ODE_Q}
\end{equation}
which has the general solution
\begin{equation}
    Q(r)=\frac{\alpha Q_\mathrm{EDL}}{1-\alpha} \left[ \left( \frac{r}{r_0}\right)^2 - \left( \frac{r}{r_0}\right)^{2\alpha} \right] + Q_0 \left( \frac{r}{r_0}\right)^{2\alpha}, \label{eq:Q_norm}
\end{equation}
where $Q_\mathrm{EDL}=\pi r_0^2 \sigma_\mathrm{EDL}$ is the EDL-contributed charge. The corresponding surface charge density is then
\begin{equation}
\begin{split}
   \sigma_{\text{S}}(r) = \frac{\alpha \sigma_{\text{EDL}}}{1-\alpha} \left[1 - \alpha \left( \frac{r}{r_0} \right)^{2\alpha-2} \right] + \alpha \, \sigma_{0} \left( \frac{r}{r_0} \right)^{2\alpha-2} \label{eq:sigma_s_norm}
\end{split}
\end{equation} 
where $\sigma_{0} =  Q_0 / (\pi r_0^2)$. The model assumes a circular footprint but remains valid for arbitrary retraction dynamics (axisymmetric and non-axisymmetric), provided that each point on the surface is traversed only once by the contact line (SM Figs.~S2--S5). 

For an initially uncharged droplet $Q_{0} = 0$, several simplifications follow. First, the initial $\sigma_{\mathrm{S}}$ matches the grounded case in  \cref{eq:sigma_s_grounded}:
\begin{equation}
\begin{split} \label{eq_sigma_init}
\sigma_{\mathrm{S}}(r_{0}) = \sigma_{\mathrm{S, 0}} =  \alpha (K \varepsilon_\mathrm{L}\phi/\lambda),
\end{split}
\end{equation}
while towards the end of the evaporation, $\sigma_{\mathrm{S}}$ evolves to
\begin{equation}
\begin{split} \label{eq_sigma_final}
\sigma_{\mathrm{S}}(r \rightarrow 0) = - \sigma_{\mathrm{S, 0}} \left( \frac{\alpha}{1 - \alpha} \right) \left( \frac{r}{r_0} \right)^{2\alpha-2},
\end{split}
\end{equation}
revealing a singularity $\sigma_{\mathrm{S}}/\sigma_{\mathrm{S, 0}} \rightarrow -\infty$ as $r \rightarrow 0$. This indicates a strong buildup of surface charge with opposite polarity near the end of evaporation.

Second, \cref{eq:sigma_s_norm}  predicts polarity reversal at
\begin{equation}
\begin{split} \label{eq_r_crit}
r_\text{crit}/r_{0} &= (1/\alpha)^{1/(2\alpha - 2)},   
\end{split}
\end{equation} 
corresponding to the maximum charge
\begin{equation}
    \frac{Q_{\text{max}}}{Q_{\text{EDL}}} = \frac{\alpha}{1-\alpha} \left( \alpha^{\frac{1}{1-\alpha}} - \alpha^{\frac{\alpha}{1-\alpha}}  \right).  \label{eq_Q_max}
\end{equation}
Thus, $\alpha$ strongly governs both the reversal point and the extent of droplet charging. In the full transfer limit $\alpha \rightarrow 1$, \cref{eq_r_crit,eq_Q_max} reduce to $r_\text{crit}/r_{0} = e^{-1/2} \approx 0.6$ and $Q_{\text{max}}/Q_{\text{EDL}} = e^{-1} \approx 0.37$ [SM Fig.~S4]. 

\begin{figure*}[!htb]
\centering
\includegraphics[scale=1]{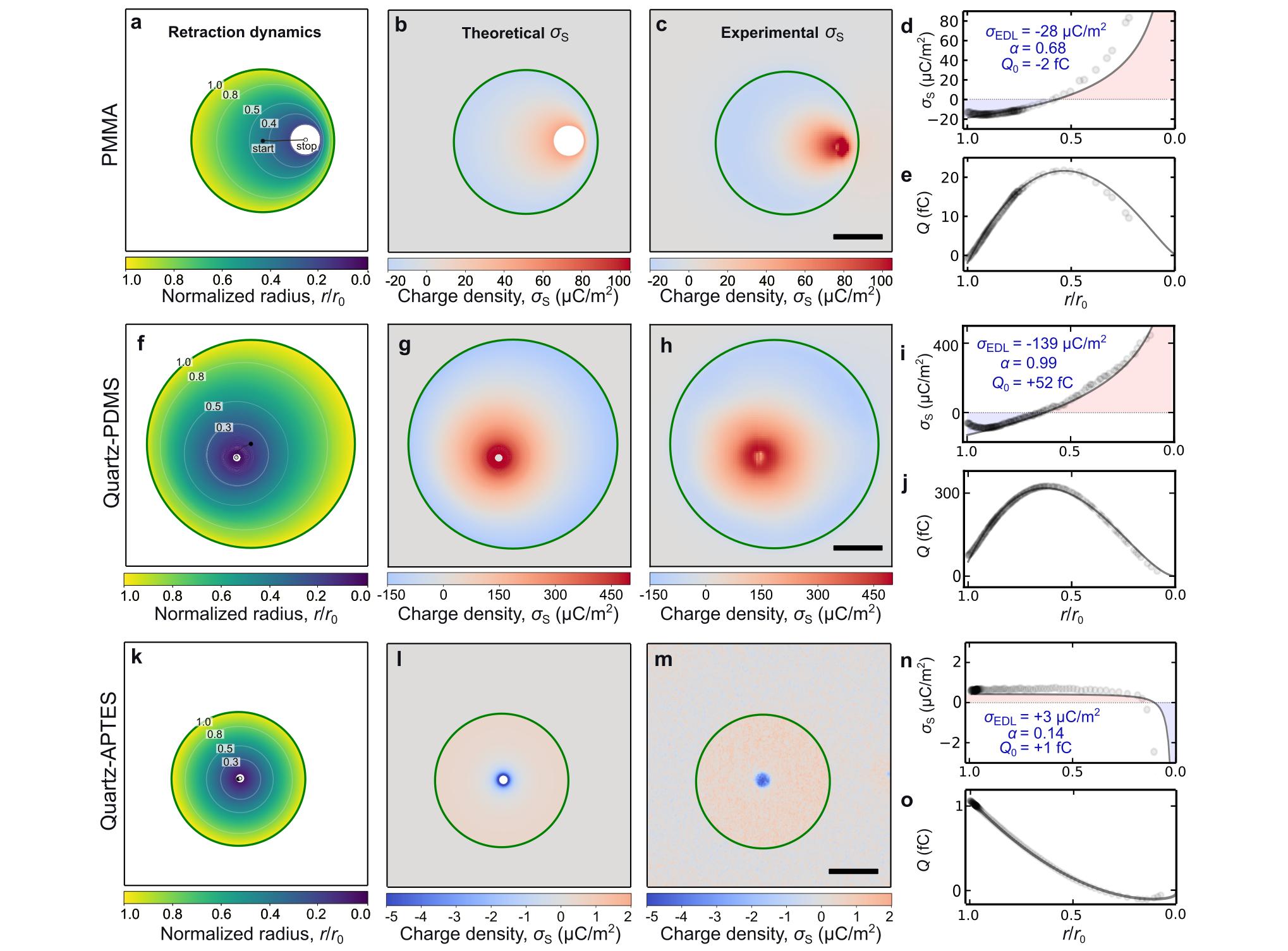}
\caption{\label{fig:bipolar_charging} (a) Retraction dynamics on PMMA, parameterized by the normalized radial coordinate $r/r_{\text{0}}$. The start and stop positions of the droplet are indicated by black and white circles, respectively. No data is available in the middle white region. (b) Theoretical $\sigma_{\text{S}}$ from \cref{eq:sigma_s_norm} vs. (c) Experimental $\sigma_{\text{S}}$ from KPFM. (d, e) $\sigma_{\text{S}}$ and $Q$ as functions of $r/r_{\text{0}}$. Lines represent theoretical predictions, while symbols denote radially averaged experimental values, whose standard deviations are smaller than the symbol size. Corresponding results for (f--j) Quartz-PDMS and (k--o) Quartz-APTES. All scale bars are \SI{20}{\micro\meter}.}
\end{figure*}

Our analytical model is validated against experiments [\cref{fig:bipolar_charging}; SM Videos S1--S3]. Using optically tracked contact-line positions $r(t)$ [\cref{fig:bipolar_charging}a, f, k], we reconstructed the surface-charge patterns via \cref{eq:sigma_s_norm} [\cref{fig:bipolar_charging}b, g, i] and compared them with experimental KPFM data [\cref{fig:bipolar_charging}c, h, m]. We chose to fit the measured charge $Q(r)$, though the spatial profile $\sigma_{\mathrm{S}}(r)$ could have been chosen. Using best-fit values for the three free parameters ($\sigma_\mathrm{EDL}$, $\alpha$, and $Q_{0}$ as indicated in \cref{fig:bipolar_charging}d, i, n), the model not only captures $Q(r)$ but also reproduces $\sigma_{\mathrm{S}}(r)$ with excellent quantitative agreement across all surfaces [\cref{fig:bipolar_charging}d, e; \cref{fig:bipolar_charging}i, j; \cref{fig:bipolar_charging}n, o]. See SM Fig.~S7 for more surfaces.

\begin{table}[!htb]
\centering
\begin{tabular}{@{\hspace{0pt}}l@{\hspace{5pt}}c@{\hspace{5pt}}c@{\hspace{5pt}}c@{\hspace{5pt}} c@{\hspace{5pt}}}
\hline
\textbf{Surface} & $\theta_{r}$ & $\alpha$ & $ \sigma_\mathrm{EDL}$           &  $\phi$ \\
                 &  ($^{\circ}$)         &          &  (\SI{}{\micro\coulomb} m$^{-2}$)&  (mV) \\
\hline
PMMA          &  $63 \pm 2$   & $0.65 \pm 0.03$ & $-28 \pm 5$    & $-8 \pm 2$  \\
Quartz-PDMS   &  $104 \pm 2$  & $0.98 \pm 0.01$ & $-170 \pm 40$     & $-49 \pm 12$ \\
Quartz-APTES  &  $65 \pm 2$   & $0.1 \pm 0.03$ & $+4 \pm 1$     & $+1 \pm 0.4$ \\ 
\hline
\end{tabular}
\caption{Fitted values of $\alpha$ and $\sigma_\mathrm{EDL}$. Errors are standard deviations for multiple droplets. $\phi$ is calculated from $\sigma_\mathrm{EDL}$ assuming $\lambda =$ 200 nm and $K$ = 1. See SM Table S1, 2}
\label{tab:fitted_values}
\end{table}

\emph{Discussion.} Our model quantitatively captures surface charging across different substrates (PMMA and Quartz) and surface chemistries (PDMS and APTES). In all cases, it reproduces key features of the charging process, including polarity reversal and the strong amplification of surface charge near the final stages of evaporation [\cref{eq_sigma_final}]. Notably, substantial charging is observed even on hydrophilic surfaces, particularly PMMA [\cref{fig:bipolar_charging}a--e]. Bipolar patterns also emerge from evaporating condensate drops, underscoring the generality of the phenomenon [SM Fig.~S6]. 

Importantly, fitting the model to experiments allows us to extract key parameters of the EDL: its charge density $\sigma_{\text{EDL}}$ and charge transfer coefficient $\alpha$ [\cref{tab:fitted_values}; see SM Table S1, 2 for extended data on other surfaces]. Together, the parameters govern surface charging: $\sigma_{\text{EDL}}$ sets the polarity and magnitude of the initial charge deposition [\cref{eq_sigma_init}], while $\alpha$ controls the point of polarity reversal and the extent of droplet charging [\cref{eq_r_crit,eq_Q_max}]. 

Variations in $\sigma_{\text{EDL}}$ explain the observed differences in $\sigma_{\mathrm{S}}(r_{0})$ across surfaces, ranging from \SI{-104}{\micro\coulomb} m$^{-2}$ for Quartz-PDMS to +\SI{0.5}{\micro\coulomb} m$^{-2}$ for Quartz-APTES [\cref{fig:bipolar_charging}d, i, n]. Surface potential values estimated via $\phi \approx \sigma_{\text{EDL}} \lambda /\varepsilon_{\text{L}}$ are broadly consistent with literature zeta-potential data \cite{Kirby.2004}, with discrepancies attributable to linearization in the model and the assumption of constant $\phi$. Likewise, differences in $\alpha$ account for the variation in reversal point, with $r_{\text{crit}}/r_{0}$ ranging from 0.6 for Quartz-PDMS to 0.2 for Quartz-APTES. 

Interestingly, Quartz-APTES exhibits a markedly low $\alpha = 0.1$, while PMMA shows an intermediate value of  $\alpha = 0.66$, and Quartz-PDMS approaches the full-transfer limit with $\alpha = 0.98$. These differences suggest distinct underlying charge-transfer mechanisms. On Quartz-APTES, charging occurs via amine groups, which act as Lewis bases that readily protonate in water. In contrast, the siloxane backbone of PDMS likely facilitates charging through weaker, physical adsorption of hydroxide ions.  These results highlight the critical role of chemical functionality---and possibly the identity of the charge carriers (H$^{+}$ vs. OH$^{-}$ ions)---in governing surface charging behaviour. A complete mechanistic understanding, however, requires atomistic insight which is beyond the scope of this study. 
 
\emph{Conclusions.} In summary, we have demonstrated that evaporating water droplets result in bipolar surface charging on both hydrophobic and hydrophilic surfaces. Our analytical model quantitatively captures the charging process and reveals the critical influence of surface chemistry. The observed variations in charging behaviour suggest dinstinct charge-transfer mechanisms for different functional groups. Gaining deeper insight into these mechanisms will enable more precise control of surface charging for applications ranging from electrostatic patterning to improved functional coatings.

\emph{Acknowledgements.} We thank H.-J. Butt for inspiring discussions. A.D.R is supported by European Union’s Horizon 2020 research and innovation program (grant agreement no. 883631). D.D. acknowledges support from KAUST and OIST start-up funds.  

D.D. proposed the work and supervised the research; N.S. designed and conducted the experiments; A.D.R. developed the analytical model; N.A. prepared the Quartz substrate and gold nanostrips; A.D.R., D.D., and N.S. jointly interpreted the results and prepared the manuscript.

\end{document}